\documentclass[11pt]{article}

\usepackage[a4paper]{geometry}
\usepackage[parfill]{parskip}    
\usepackage[usenames,dvipsnames]{xcolor}
\usepackage[english]{babel}
\usepackage{cite,enumerate,enumitem,booktabs,float,graphicx}

\makeatletter
\g@addto@macro\bfseries{\boldmath}
\makeatother

\usepackage{amsmath,amssymb,bm,mathtools,tensor}
\numberwithin{equation}{section}

\usepackage{authblk}

\usepackage[pdftex]{hyperref}
\definecolor{dark-blue}{rgb}{0.15,0.15,0.4}
\hypersetup{
  colorlinks,
  linkcolor={dark-blue},
  citecolor={blue}
}

\newcommand{\beq}{\begin{equation}}
\newcommand{\eeq}{\end{equation}}

\DeclareMathOperator{\Tr}{Tr}

\newcommand{\scA}{\mathcal{A}}

\newcommand{\scM}{\mathcal{M}}
\newcommand{\scV}{\mathcal{V}}
\newcommand{\scAhat}{\hat{\mathcal{A}}}

\newcommand{\scMhat}{\hat{\mathcal{M}}}
\newcommand{\scVhat}{\hat{\mathcal{V}}}

\newcommand{\CC}{\mathbb{C}}

\newcommand{\RR}{\mathbb{R}}

\renewcommand{\tt}{\mathfrak{t}}
\newcommand{\SL}{\mathrm{SL}}
\renewcommand{\sl}{\mathfrak{sl}}

\newcommand{\so}{\mathfrak{so}}

\newcommand{\upd}{\mathrm{d}}
\newcommand{\upE}{{\mathsf{E}}}

\title{\textbf{$\SL(2,\RR)$ families of Kerr black holes}}
\date{}

\author[]{Robert Penna\footnote{pennar@sunypoly.edu}}
\affil[]{
  Department of Mathematics and Physics \\
  SUNY Polytechnic Institute,
  Utica, NY 13502 USA
}

\begin{document}

\maketitle

\thispagestyle{empty}

\begin{abstract}

The stationary, axisymmetric sector of vacuum general relativity (with zero cosmological constant) enjoys an $\SL(2,\RR)$ symmetry called the Matzner-Misner group.  
We study the action of the Matzner-Misner group on the Kerr black hole.  We show that the group acts naturally on a three parameter generalization of the usual two parameter Kerr solution.  The new parameter represents a large diffeomorphism which gives the spacetime an asymptotic angular velocity.    
We explain how the $\SL(2,\RR)$ symmetry organizes the space of three parameter Kerr solutions into the classical analogue of principal series representations.    
We show that the $\SL(2,\RR)$ Casimir operator is the Bekenstein-Hawking entropy.  
The Matzner-Misner group sits inside a much larger Kac-Moody symmetry called the Geroch group.  
We show that the Kac-Moody level of the Kerr black hole is the Bekenstein-Hawking entropy.

\end{abstract}

\newpage

\section{Introduction}

The Geroch group is a well-known Kac-Moody symmetry of the stationary, axisymmetric sector\footnote{Actually, there is an avatar of the Geroch group for every sector of general relativity with two commuting Killing vectors.  We focus on the stationary, axisymmetric sector because it is the one that contains black holes.} of general relativity \cite{geroch1971method,Julia:1981wc,breitenlohner2005explicit,Breitenlohner:1986um,Nicolai:1991tt,nicolai1992hyperbolic}.
It is an important tool for generating exact solutions.  

In the present work, we compute Noether charges for the Geroch group and we use these charges to classify and organize the space of classical configurations.  This is a familiar program in quantum mechanics: when a group $G$ is a symmetry of a quantum system, the states can be organized into representations of $G$.  This story has an analogue in the classical world: when a group $G$ is a symmetry of a classical system, the classical configurations can be organized into coadjoint orbits of $G$.  Coadjoint orbits are symplectic manifolds, the classical analogue of Hilbert spaces \cite{vergne1983representations,witten1988coadjoint,kirillov2025lectures}.

We mostly focus on an $\SL(2,\RR)$ subgroup of the Geroch group called the Matzner-Misner group and we study its action on the Kerr black hole.  The main surprise is this: the Matzner-Misner group acts not on the usual two parameter Kerr solution but on a three parameter generalization thereof.  The additional parameter represents a large diffeomorphism.  It can be thought of as giving the spacetime an overall angular velocity.  Large diffeomorphisms such as this have been a subject of major interest in recent years, as they might be measurable with gravitational wave detectors using the gravitational memory effect \cite{grishchuk1989gravitational,strominger2018lectures,bieri2024gravitational}.  An interesting question for the future is to understand if it is possible to design an experiment that can measure the extra parameter in the Kerr solution described herein.  This new parameter is really just the first in a tower of parameters that are generated by the action of the Geroch group.  There is potentially a lot of work to be done exploring this tower of parameters and understanding which ones are physically measurable.  A more speculative direction is to explore whether this tower of parameters is somehow related to the problem of counting Kerr black hole microstates.

Assigning Noether charges to Geroch group transformations is not straightforward.  The main technical novelty of the present work is that we provide a new prescription for computing these charges.  Here is the problem: the Geroch group is most naturally described as a symmetry of dimensionally reduced general relativity.  We are considering the stationary, axisymmetric sector, so we want to reduce with respect to time, $t$, and an angular coordinate, $\varphi$.  The problem is that when we integrate over time, the reduced action picks up an overall factor of $\infty$.  This seems to make the Noether charges ill-defined.  (In the solution generating literature this problem is not so important because the emphasis is on the equations of motion.  The normalization of the action is not too important in that case.)

Our solution (following the work of Gibbons and Hawking \cite{gibbons1977action} on black hole thermodynamics) is to Wick rotate to imaginary time and integrate over the thermal circle.  Now the dimensionally reduced action is well defined.  It picks up an overall factor of $\beta$, the inverse temperature, from the integral over imaginary time.  The action of the Geroch group is still well defined and, best of all, the Noether currents are finite and well defined.  To turn the Noether currents into Noether charges, we need to choose some prescription for integrating them.  Our choice is to integrate them over the event horizon.

Now that we have the ability to assign finite Noether charges to Geroch transformations, we can just start computing and see what we find.  It is simplest to start by looking at the Matzner-Misner group.  The Noether current, $j$, is valued in the dual of the Lie algebra, $\sl(2,\RR)^*$.  The action of the Matzner-Misner group on $\sl(2,\RR)^*$ foliates the latter into families of coadjoint orbits.  We show that the three parameter Kerr solutions fall on a family of coadjoint orbits that can be described as one-sheeted hyperboloids.  These are well-known to be the classical analogue of principal series representations.  We show that the Casimir operator, which controls the overall size of the hyperboloid, is nothing other than the Bekenstein-Hawking entropy.  All three parameters in the generalized Kerr solution play an essential role in this story.  The overall picture is summarized in Figure \ref{fig:kerr}.  This figure shows one such hyperboloid and some solid curves which represent black holes with the same mass and spin (the new, third parameter is varying along the curves).  Readers who are pressed for time are urged to just look at this figure as it more or less summarizes our main results about the Matzner-Misner group.

In the last section of this work, we turn to the full Geroch group and compute the Kac-Moody level of the Kerr black hole.  This requires a digression into the action of the Ehlers group.  The Ehlers group is another $\SL(2,\RR)$ subgroup of the Geroch group.  The action of the Ehlers group does not commute with the action of the Matzner-Misner group and, in fact, together these two groups generate the Geroch group.  We compute the Noether charges for the Ehlers group.  Then, by comparing these charges with the charges coming from the Matzner-Misner group, we can read off the Kac-Moody level.  It is given by the Bekenstein-Hawking entropy.  We expect this result to be an important ingredient in future efforts to classify how the tower of parameters generated by the Geroch group is organized into Kac-Moody coadjoint orbits.

\section{Matzner-Misner symmetry}

The Einstein-Hilbert action on a manifold $M$ of Euclidean signature is (setting $G=1$)
\beq
I_{\mathrm{EH}} = -\frac{1}{16\pi} \int_M \upd^4 x \sqrt{g} R \,.
\eeq
The field equation is $R_{\mu\nu} = 0$.  Any stationary, axisymmetric solution can be brought to the canonical Weyl form \cite{stephani2009exact}
\beq\label{eq:weyl}
\upd s^2 = e^{-2U} \left[ e^{2\gamma} \left( \upd \rho^2 + \upd z^2 \right) + \phi^2 \upd \varphi^2  \right] 
		+ e^{2U} \left( \upd t_\upE - i A \upd \varphi \right)^2 \,.
\eeq
Real and imaginary time are related by $t=-i t_\upE$ and we take $t_\upE$ to be a periodic variable with period $\beta$.  
The metric is complex (unless $A=0$, as for the Schwarzschild black hole, in which case it is real).  All of the functions in the metric ($U$, $A$, $\phi$, and $\gamma$) are real.
The assumption that the solution is stationary and axisymmetric means the functions in the metric depend on $z$ and $\rho$ only.  
Reinserting \eqref{eq:weyl} into the Einstein-Hilbert action and dropping total derivatives gives
\beq\label{eq:I2d}
I_{\mathrm{2d}} = \frac{\beta}{4} \int_\Sigma \upd^2 x 
	\left[ \phi (\partial U)^2
		- \frac{1}{4} \phi^{-1} e^{4U} (\partial A)^2
		- \partial \phi \cdot \partial \gamma \right] .
\eeq
We picked up a factor of $2\pi \beta$ from the integrals over $t_{\upE}$ and $\varphi$.  The reduced action is defined on a two dimensional manifold, $\Sigma$, with coordinates $z$ and $\rho$. The metric on $\Sigma$ is the flat metric\footnote{We could try to make life more interesting by coupling $I_{\rm 2d}$ to 2d gravity.  But it is always possible to make the 2d metric conformally flat using a 2d diffeomorphism and then the conformal factor can be absorbed into $\gamma$.  In fact, it is well known that the 2d Einstein equation we would obtain is equivalent to the equations of motion coming from \eqref{eq:I2d}.    (See \cite{Breitenlohner:1986um} for details.) So nothing is really lost by sticking with a flat metric on $\Sigma$.}.

The equations of motion are
\begin{align}
\partial \cdot \left( \phi \partial U \right) + \frac{1}{2} \phi^{-1} e^{4U} (\partial A)^2 	&=0 \,, \label{eq:eom1} \\
\partial \cdot \left( \phi^{-1} e^{4U} \partial A \right)							&=0 \,, \label{eq:eom2} \\
\partial^2 \gamma + (\partial U)^2 + \frac{1}{4} \phi^{-2} e^{4U} (\partial A)^2 		&=0 \,, \label{eq:eom3} \\
											\partial^2 \phi 			&=0 \,. \label{eq:eom4}
\end{align}

The action \eqref{eq:I2d} has two shift symmetries, under $\gamma \rightarrow \gamma + 1$ and $A \rightarrow A + 1$.  
We will return to the $\gamma \rightarrow \gamma+1$ symmetry later (it gives the Kac-Moody level).  
The $A\rightarrow A+1$ symmetry forms part of an $\SL(2,\RR)$ symmetry called the Matzner-Misner group. 
To draw out the full Matzner-Misner symmetry, we introduce a matrix-valued field,
\beq
\scV \equiv
\begin{pmatrix}
e^{-U} \phi^{1/2} 	&&	0	\\
-iA e^U \phi^{-1/2}	&&	 e^U \phi^{-1/2}	
\end{pmatrix} .
\eeq
Further define $\scM \equiv \scV^T \scV$ and $\scA \equiv \scM^{-1} d\scM$.  The action becomes
\beq
I_{\mathrm{2d}} = \frac{\beta}{4} \int_\Sigma \upd^2 x 
	\left[ \frac{1}{8} \phi \Tr \scA^2 
		+ \partial \phi \cdot \partial U
		- \frac{1}{4} \phi^{-1} (\partial \phi)^2
		- \partial \phi \cdot \partial \gamma \right] .
\eeq

Observe that $\scV$ is a lower triangular matrix in $\SL(2,\CC)$ and consider the following action of $\epsilon\in \sl(2,\RR)$ on $\scV$:
\beq\label{eq:mmaction}
\delta_\epsilon \scV = -h_\epsilon(z,\rho) \scV + \scV \epsilon \,.
\eeq
In this formula, $h_\epsilon(z,\rho)$ is a spatially varying element of $\so(2)$.  It is a compensating gauge transformation which depends on $\epsilon$.  Its job is to restore the lower triangular form of $\scV$.

A convenient basis for $\sl(2,\RR)$ is
\beq
\tt_1 = \begin{pmatrix}
0	&&	-i	\\
-i	&&	0	
\end{pmatrix}	, \qquad
\tt_2 = \begin{pmatrix}
0	&&	-i	\\
i	&&	0	
\end{pmatrix}	, \qquad
\tt_3 = \begin{pmatrix}
1	&&	0	\\
0	&&	-1	
\end{pmatrix}	.
\eeq
This is a somewhat unconventional basis but we choose it because the factors of $i$ will cancel against the factor of $i$ in $\scV$ and lead to real-valued Noether charges.  
The commutation relations are
\beq
[ \tt_3 , \tt_1 ] = 2 \tt_2 \,, \qquad
[ \tt_3 , \tt_2 ] = 2 \tt_1 \,, \qquad
[ \tt_1 , \tt_2 ] = 2 \tt_3 \,.
\eeq
The metric on $\sl(2,\RR)\approx \RR^{1,2}$  is the Minkowski metric and the timelike direction is $\tt_1$.

Plugging the $\tt_a$ into \eqref{eq:mmaction}, we find the field transformations
\begin{alignat}{2}
\delta_{\tt_1} U &= -A		\,, \qquad		&&\delta_{\tt_1} A = 1 + e^{-4U} \phi^2 + A^2 \,, 	\\
\delta_{\tt_2} U &= -A		\,, \qquad 		&&\delta_{\tt_2} A = -1 + e^{-4U} \phi^2 + A^2 \,, \\
\delta_{\tt_3} U &= -1	\,, 		\qquad 		&&\delta_{\tt_3} A = 2 A \,.
\end{alignat}
The shift symmetry $A\rightarrow A+1$ we remarked upon earlier comes from the action of $\frac{1}{2} ( \tt_1 - \tt_2 )$.
The above transformations are symmetries of the sigma model term, $\Tr \scA^2$, but they are not symmetries of the full action.  To get symmetries of the full action, we further impose
\begin{alignat}{2}
\delta_{\tt_1} \phi &= 0	\,, \qquad		&&\delta_{\tt_1} \gamma = -A \,, 	\label{eq:mmgamma1}	\\
\delta_{\tt_2} \phi &= 0	\,, \qquad 		&&\delta_{\tt_2} \gamma = -A \,,	\label{eq:mmgamma2}	 \\
\delta_{\tt_3} \phi &= 0 	\,, \qquad 		&&\delta_{\tt_3} \gamma = -1 \,.		\label{eq:mmgamma3}
\end{alignat}
Actually, on the last line, we could set $\delta_{\tt_3} \gamma = 0$ and still have a symmetry.  We will explain the reason for including this transformation in a moment.  We comment in passing that it is rather remarkable that it is possible to extend the symmetry to the full action by making these extra prescriptions.

The Noether currents are
\begin{align}
j_1	&=  \frac{\beta}{4} \left[ -2 \phi A \partial U
				- \frac{1}{2} \phi^{-1} e^{4U} \left( 1 + e^{-4U} \phi^2 +A^2 \right) \partial A
				+ A \partial \phi  \right] , 	\label{eq:noether1} \\
j_2	&=  \frac{\beta}{4} \left[ -2 \phi A \partial U
				- \frac{1}{2} \phi^{-1} e^{4U} \left( -1 + e^{-4U} \phi^2 + A^2 \right) \partial A
				+ A \partial \phi \right] , 	\label{eq:noether2} \\	
j_3	&=  \frac{\beta}{4} \left[ -2 \phi \partial U
				- \phi^{-1} e^{4U} A \partial A 
				+ \partial \phi \right] .		\label{eq:noether3}
\end{align}
The conservation laws for these currents follow from the equations of motion \eqref{eq:eom1}--\eqref{eq:eom4}.   A more compact 
expression for the currents is
\beq\label{eq:jmm}
j = \frac{\beta}{8} \phi \scA \,.
\eeq
The $\sl(2,\RR)$ components are $j_a = \Tr(j \tt_a)$.  
From \eqref{eq:mmaction} and the definition of $\scA$, we see that the action of the Matzner-Misner group on the current is $j \rightarrow g^{-1} j g$.  This is the coadjoint action of $\SL(2,\RR)$ on the dual of its Lie algebra.  The reason we included $\delta_{\tt_3} \gamma = -1$ in \eqref{eq:mmgamma3} is that it leads to the correct $\SL(2,\RR)$ transformation law for the Noether currents.

\subsection{The Kerr black hole}

Now that we have the Noether currents \eqref{eq:noether1}--\eqref{eq:noether2}, we can evaluate them on the Kerr black hole.  
The first step is to recall the Kerr solution \cite{stephani2009exact}:
\begin{align}
e^{2U} &= \frac{p^2 x^2 + q^2 y^2 -1}{(px+1)^2 + q^2 y^2} \,, \qquad
e^{2\gamma} = \frac{p^2 x^2 + q^2 y^2 -1}{p^2 (x^2 - y^2)} \,, \\
A &= \frac{2mq}{p^2 x^2 + q^2 y^2 -1}(1-y^2) (px + 1) + b \,, \qquad \phi = \rho \,.
\end{align}
We have written the solution in prolate spheroidal coordinates ($x$ and $y$) because it is unfortunately difficult to write the solution directly in Weyl coordinates (with the exception of $\phi = \rho$).  We will explain the relationship to Weyl coordinates below.  
The solution as we have written it has three independent parameters.  The first two are the usual mass and spin parameters, $m$ and $a$.  They are related to $p$ and $q$ by
\beq
q= a/m \,, \qquad p = \sqrt{1-q^2} \,.
\eeq
The third parameter, $b$, only appears in $A$.  It is usually set to zero.  We include it because it is generated by the action of the Matzner-Misner symmetry $A\rightarrow A+b$ discussed above.  For lack of a better name, we will call $b$ the shift parameter.  We can view this parameter as a large diffeomorphism.  Indeed, the Kerr metric in Lorentzian signature has the form
\beq
\upd s^2 = e^{-2U} \left[ e^{2\gamma} \left( \upd \rho^2 + \upd z^2 \right) + \phi^2 \upd \varphi^2  \right] 
		- e^{2U} \left( \upd t + A \upd \varphi \right)^2 \,.
\eeq
So the shift $A\rightarrow A+b$ is equivalent to the coordinate change $t \rightarrow t + b \varphi$.  Since $e^{2U} \rightarrow 1$ at the asymptotic boundary, this is a large diffeomorphism.  These diffeomorphisms are physical symmetries; the gravitational memory effect is one way to measure them \cite{strominger2018lectures}.  One of the messages of the present work is that this particular large diffeomorphism is special because of its relationship to the Matzner-Misner group. 

\begin{figure}
\centering
\includegraphics[width=0.2\linewidth]{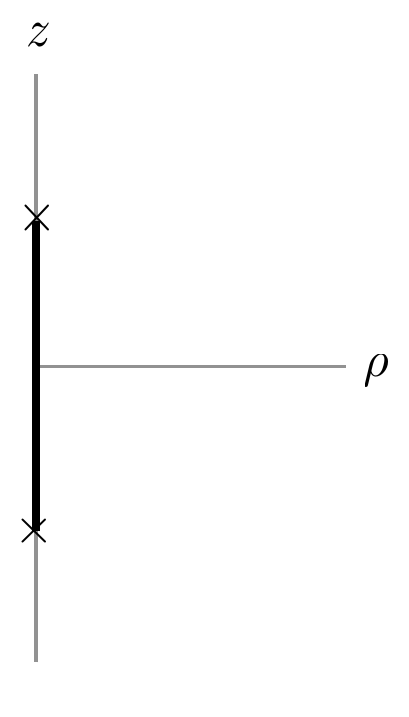}
\caption{In Weyl coordinates, the event horizon sits at $\rho = 0 $ and the $z$ coordinate runs from $-\sigma$ to $+\sigma$.}
\label{fig:horizon}
\end{figure}

Returning to the Kerr solution, we need to recall how prolate spheroidal coordinates ($x$ and $y$) are related to Weyl coordinates ($\rho$ and $z$).  Actually, it is difficult to state this relationship directly.  It is more natural to state how each set of coordinates is related to Boyer-Lindquist coordinates ($r$ and $\theta$), and so obtain an implicit relationship between prolate spheroidal and Weyl coordinates.  The former are related to Boyer-Lindquist coordinates by
\beq
\sigma x = r - m \,, \qquad
y = \cos\theta \,,
\eeq
where $\sigma \equiv \sqrt{m^2 - a^2}$.  The latter are related to Boyer-Lindquist coordinates by
\beq
\rho = \sqrt{r^2 - 2mr + a^2} \sin\theta \,, \qquad
z = (r- m ) \cos\theta \,.
\eeq
The event horizon is at $r_+ \equiv m + \sigma$.  Observe that $\rho = 0$ at the event horizon and $z=\sigma \cos\theta$.  The $\theta$ coordinate runs from $0$ to $\pi$, so the $z$ coordinate runs from $-\sigma$ to $+\sigma$.  We summarize these remarks in Figure \ref{fig:horizon}.  

To turn the Noether currents \eqref{eq:noether1}--\eqref{eq:noether2} into Noether charges, we need a prescription for integrating the currents.  Our prescription is to integrate the currents over the event horizon.  Thus, we define the charge to be
\beq\label{eq:charge}
q \equiv \int_{-\sigma}^\sigma \upd z \, j^\rho \bigg\vert_{\rho = 0} \,.
\eeq
To gain some feeling for this definition, let us first look at something simpler than the Matzner-Misner currents.  The action \eqref{eq:I2d} has a shift symmetry $\gamma \rightarrow \gamma - 1$, as observed earlier.  The Noether current for this symmetry is
\beq
J = \frac{\beta}{4} \partial \phi \,.
\eeq
So the charge formula gives
\beq
Q = \int_{-\sigma}^\sigma \upd z \, J^\rho \bigg\vert_{\rho = 0}
	= \frac{\beta}{4} \int_{-\sigma}^\sigma \upd z 
	= \frac{\beta \sigma}{2} \,.
\eeq
The Kerr black hole has $\beta = 4 \pi m r_+/\sigma$, thus
\beq
Q = 2\pi m r_+ = \frac{A_H}{4} \,, 
\eeq
where $A_H \equiv 8 \pi m r_+$ is the surface area of the event horizon.  So $Q$ is nothing other than the Bekenstein-Hawking entropy of the black hole.  
As discussed in the Introduction, our motivation for computing these charges in the first place is that we would like them to provide a tool for organizing the space of generalized Kerr solutions (generalized to include shift parameters like $b$).  
The simplicity of $Q$ suggests our prescription \eqref{eq:charge} is a sensible choice.

We have seen above that the Matzner-Misner current, $j$, transforms as $j\rightarrow g^{-1} j g$ under the action of the Matzner-Misner group.  This action has a well-known invariant, the Casimir function,
\beq
R^2 \equiv -q_1^2 + q_2^2 + q_3^2 \,.
\eeq
Computing this function for the Kerr metric using \eqref{eq:charge} gives
\beq
R = \frac{A_H}{4}  \,.
\eeq
So the Casimir function is also given by the Bekenstein-Hawking entropy.

The current, $j$, is valued in $\sl(2,\RR)^*$, the dual of the Lie algebra\footnote{Actually, these spaces are isomorphic in the case of $\SL(2,\RR)$ so the distinction between the Lie algebra and its dual is not too important.  We will maintain the distinction to remain consistent with the theory for general Lie groups.} $\sl(2,\RR)$.  The charges, $q_1$, $q_2$, and $q_3$, provide a natural set of coordinates on $\sl(2,\RR)^*$.  The Killing metric endows $\sl(2,\RR)^*$ with the structure of three dimensional Minkowski space: $\sl(2,\RR)^* \approx \RR^{1,2}$.  The time-like direction is $q_1$.  

The action of the Matzner-Misner group, $q \rightarrow g^{-1} q g$, foliates $\sl(2,\RR)^* \approx \RR^{1,2}$ into families of symplectic manifolds.  These manifolds are the coadjoint orbits of the Matzner-Misner group.  Quantization turns coadjoint orbits into irreducible representations.  So coadjoint orbits are the classical analogue of irreducible representations.   The coadjoint orbits of $\SL(2,\RR)$ are well-known.  They fall into several families: a family of one-sheeted hyperboloids, a family of two-sheeted hyperboloids, the null cone, and the origin.  We refer to \cite{vergne1983representations,witten1988coadjoint,kirillov2025lectures} for reviews.

\begin{figure}
\centering
\includegraphics[width=0.6\linewidth]{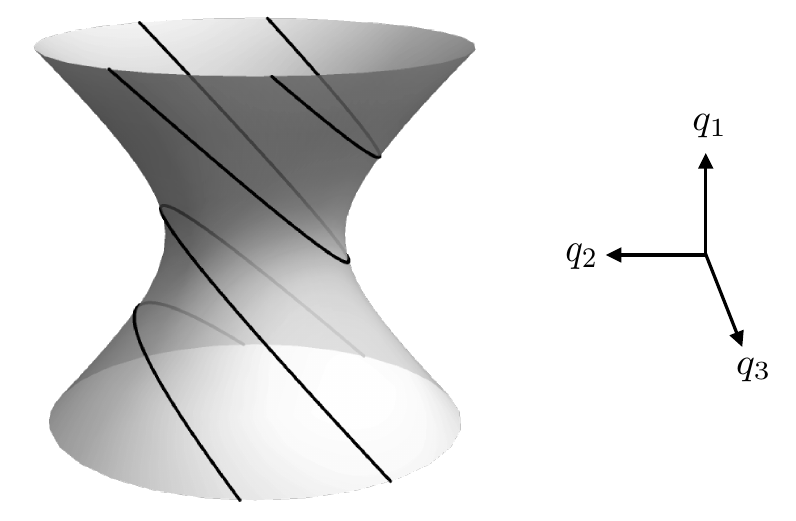}
\caption{The three parameter Kerr solutions sit on a family of one-sheeted hyperboloids in $\sl(2,\RR)^* \approx \RR^{1,2}$.  The overall scale of the hyperboloid is set by the Bekenstein-Hawking entropy.  Each solid black curve is a family of Kerr solutions with fixed $m$ and $a$ and varying $b$.  In this plot, the spin parameters of the four solid black curves are $a = -0.999$,  $-0.7$, $0.7$, and $0.999$ and the masses are $m = 1.05$, $0.93$, $0.93$, and $1.05$ (moving diagonally upward from bottom left to top right).  The shift parameter, $b$, varies along the solid black curves.}
\label{fig:kerr}
\end{figure}

Which kind of coadjoint orbit corresponds to the Kerr solution?  Since $R^2 >0$, it must be that the Kerr solutions fall on the one-sheeted hyperboloids.   We can see this very directly by computing the $q$'s using eqn. \eqref{eq:charge} and drawing a picture: see Figure \ref{fig:kerr}.   The Kerr solutions fall on the one-sheeted hyperboloids.  These are the classical analogue of principal series representations. The Casimir function (alias the Bekenstein-Hawking entropy), $R=A_H/4$, sets the overall scale of the hyperboloid.  Kerr solutions with the same Bekenstein-Hawking entropy are represented by points on the same hyperboloid in $\RR^{1,2}$.  In Figure \ref{fig:kerr}, we have drawn some solid black curves: each curve represents a one-parameter family of Kerr solutions with fixed $m$ and $a$ and varying shift parameter, $b$.  We see that it really is essential to include the shift parameter in order to get a full picture of how the $\SL(2,\RR)$ symmetry organizes the space of Kerr solutions.

We remark that there are two equivalent ways to arrive at Figure \ref{fig:kerr}.  One is simply to plot $(q_1,q_2,q_3) \in \RR^{1,2}$ for a family of Kerr solutions with fixed Bekenstein-Hawking entropy.  Another is to plot the solutions with $b=0$, and then act with $q\rightarrow g^{-1} q g$, where
\beq
g = \begin{pmatrix}
1	&&	0	\\
-ib	&&	1 
\end{pmatrix} .
\eeq
The action of this matrix sweeps out the one-sheeted hyperboloids and the black curves in Figure \ref{fig:kerr}.

The action of the full Geroch group generates an infinite tower of new parameters.  It would be interesting to understand the analogue of Figure \ref{fig:kerr} for this tower of parameters and to design experiments to measure them.    

\section{Kac-Moody level}

The Geroch group is generated by the Matzner-Misner group and a second $\SL(2,\RR)$ subgroup called the Ehlers group.  To compute the Kac-Moody level of the Kerr black hole, we need to make a digression into the action of the Ehlers group.  The Ehlers group is a hidden symmetry of the 2d action, $I_{\mathrm{2d}}$.  To bring this symmetry to light, we begin by taking a closer look at the $A$ equation of motion \eqref{eq:eom2}:
\beq
\partial_z \left( \phi^{-1} e^{4U} \partial_z A \right) + \partial_\rho \left( \phi^{-1} e^{4U} \partial_\rho A \right) = 0 \,.
\eeq
This equation would become automatic if we could find a dual potential\footnote{We will not use this in what follows but we note for completeness that the Kerr solution has $\hat{A} = 2qy/[(px+1)^2 + q^2 y^2] + c$.  The new parameter, $c$, is shifted by the action of the Ehlers group.  }, $\hat{A}$, satisfying
\begin{align}
\partial_z A 	&= 	\phi e^{-4U} \partial_\rho \hat{A} \,, 	\label{eq:Ahat1}	 \\
\partial_\rho A	&= - \phi e^{-4U} \partial_z \hat{A} \,. 		\label{eq:Ahat2} 
\end{align}
The consistency condition for the existence of $\hat{A}$ is $\partial_\rho \partial_z A  - \partial_z \partial_\rho A = 0$.  This can be interpreted as an equation of motion for $\hat{A}$:
\beq
\partial \cdot \left( \phi e^{-4U} \partial \hat{A} \right) = 0 \,.
\eeq
Now we return to the complete set of equations of motion \eqref{eq:eom1}--\eqref{eq:eom2}, and we use eqns. \eqref{eq:Ahat1}--\eqref{eq:Ahat2} to get an alternative set with $\hat{A}$ in place of $A$:
\begin{align}
\partial \cdot \left( \phi \partial U \right) + \frac{1}{2} \phi e^{-4U} (\partial \hat{A})^2 	&=0 \,, \label{eq:eom1hat} \\
\partial \cdot \left( \phi e^{-4U} \partial \hat{A} \right) 							&=0 \,, \label{eq:eom2hat} \\
\partial^2 \gamma + (\partial U)^2 + \frac{1}{4} e^{-4U} (\partial \hat{A})^2 			&=0 \,, \label{eq:eom3hat} \\
											\partial^2 \phi 			&=0 \,. \label{eq:eom4hat}
\end{align}
These equations of motion come from the action
\beq\label{eq:I2dhat}
\hat{I}_{\mathrm{2d}} = \frac{\beta}{4} \int_\Sigma \upd^2 x 
	\left[ \phi (\partial U)^2
		+ \frac{1}{4} \phi e^{-4U} (\partial \hat{A})^2
		- \partial \phi \cdot \partial \gamma \right] .
\eeq
This action has an $\SL(2,\RR)$ symmetry called the Ehlers group which is distinct from the Matzner-Misner group discussed in the previous section.  One of the new symmetries is obvious: $\hat{A} \rightarrow \hat{A} + 1$ is a symmetry of $\hat{I}_{\mathrm{2d}}$.  To make the full symmetry explicit, it is again helpful to introduce an $\SL(2,\CC)$-valued field.  This time we choose
\beq
\hat{\scV} \equiv 
\begin{pmatrix}
e^U			&&	0	\\
-i e^{-U} \hat{A}	&&	e^{-U} 
\end{pmatrix} .
\eeq
Let 
\beq
\eta \equiv 
\begin{pmatrix}
1	&&	0	\\
0	&&	-1	
\end{pmatrix} .
\eeq
In parallel with our previous discussion, we define $\scMhat \equiv \scVhat^T \eta \scVhat$ and $\scAhat \equiv \scMhat^{-1} d\scMhat$. 
In terms of these variables, the action becomes
\beq\label{eq:I2dhat2}
\hat{I}_{\mathrm{2d}} = \frac{\beta}{4} \int_\Sigma \upd^2 x 
	\left[ \frac{1}{8} \phi \Tr \scAhat^2
		- \partial \phi \cdot \partial \gamma \right] .
\eeq

Use the same $\sl(2,\RR)$ basis as before:
\beq
\hat{\tt}_1 = \begin{pmatrix}
0	&&	-i	\\
-i	&&	0	
\end{pmatrix}	, \qquad
\hat{\tt}_2 = \begin{pmatrix}
0	&&	-i	\\
i	&&	0	
\end{pmatrix}	, \qquad
\hat{\tt}_3 = \begin{pmatrix}
1	&&	0	\\
0	&&	-1	
\end{pmatrix}	.
\eeq
The transformation rule is
\beq\label{eq:ehlersaction}
\delta_\epsilon \scVhat = -h(x) \scVhat + \scVhat \epsilon \,,
\eeq
except now $h(x) \in \so(1,1)$.  
Arrive at 
\begin{alignat}{2}
\delta_{\hat{\tt}_1} U &= \hat{A}	\,, \qquad		&&\delta_{\hat{\tt}_1} \hat{A} =  1 - e^{4U} + \hat{A}^2 \,, 	\\
\delta_{\hat{\tt}_2} U &= \hat{A}	\,, \qquad 		&&\delta_{\hat{\tt}_2} \hat{A} =  -1 - e^{4U} + \hat{A}^2 \,,	\\
\delta_{\hat{\tt}_3} U &= 1 		\,, \qquad 		&&\delta_{\hat{\tt}_3} \hat{A} = 2\hat{A}  \,.
\end{alignat}
Assume $\phi$ and $\gamma$ are inert.  Then these are symmetries of $\hat{I}_{\mathrm{2d}}$.  

The Noether currents are
\begin{align}
\hat{j}_1	&= \frac{\beta}{4} \left[ 2 \phi \hat{A} \partial U
				+ \frac{1}{2} \phi e^{-4U} \left( 1 - e^{4U} + \hat{A}^2 \right) \partial \hat{A} \right] , \\
\hat{j}_2	&= \frac{\beta}{4} \left[ 2 \phi \hat{A} \partial U
				+\frac{1}{2} \phi e^{-4U} \left( -1 - e^{4U} + \hat{A}^2 \right) \partial \hat{A} \right] , \\
\hat{j}_3	&= \frac{\beta}{4} \left[ 2 \phi \partial U
				+ \phi e^{-4U} \hat{A} \partial \hat{A} \right] .
\end{align}	
A more compact expression is
\beq
\hat{j} = \frac{\beta}{8} \phi \scAhat \,.
\eeq
The $\sl(2,\RR)$ components are $\hat{j}_a = \Tr(\hat{j} \hat{\tt}_a)$.
The current, $\hat{j}$, transforms in the coadjoint representation of $\SL(2,\RR)$.  Thus, unlike in the case of Matzner-Misner symmetry, we do not supplement the field transformations with a shift of $\gamma$.  (In the Matzner-Misner case discussed above, it was necessary to supplement the field transformations with a shift of $\gamma$ to get $j = \beta \phi \scA/8$.)

So far we have been discussing the Ehlers group as a symmetry of $\hat{I}_{\rm 2d}$.  How can we realize the Ehlers group as a symmetry of the original action, $I_{\rm 2d}$?  For that, we have to work out how the Ehlers transformations act on $A$.  The way to do this is to look again at the duality equations \eqref{eq:Ahat1}--\eqref{eq:Ahat2}:
\begin{align}
\partial_z A 	&= 	\phi e^{-4U} \partial_\rho \hat{A} \,, 		 \\
\partial_\rho A	&= - \phi e^{-4U} \partial_z \hat{A} \,. 		
\end{align}
Given the action of the Ehlers group on $U$, $\hat{A}$ and $\phi$, we can vary the duality equation to deduce the Ehlers transformations of $A$.  In particular, for the variation with respect to $\hat{\tt}_3$, we find
\begin{align}
\partial_z \delta_{\hat{\tt}_3}  A 	
			&= 	-2 \phi e^{-4U} \partial_\rho \hat{A}
			=  -2 \partial_z A   \,, 		 \\
\partial_\rho \delta_{\hat{\tt}_3} A	
			&= 2 \phi e^{-4U} \partial_z \hat{A}
			= -2 \partial_\rho  A \,. 		
\end{align}
We thus arrive at the rule\footnote{Actually, there is some ambiguity in this rule as we are free to add an additive constant to the right hand side.  Our choice to not include the constant is the standard choice in the literature.  Including the constant will change the commutation relations of the Matzner-Misner and Ehlers algebras.  It is not clear to the author what this would do to the Kac-Moody symmetry.}
\beq\label{eq:ehlersA}
\delta_{\hat{\tt}_3}  A  = - 2 A \,.
\eeq

The actions of $\hat{\tt}_1$ and $\hat{\tt}_2$ on $A$ are more complicated.  We refer to \cite{breitenlohner2005explicit,Breitenlohner:1986um,Nicolai:1991tt,nicolai1992hyperbolic} for details.

Now that have \eqref{eq:ehlersA}, we can view $\hat{\tt}_3$ as a symmetry of the original theory, $I_{\mathrm{2d}}$.  The associated Noether current is
\beq
\tilde{j}_3	=  \frac{\beta}{4} \left[ 2 \phi \partial U
				+ \phi^{-1} e^{4U} A \partial A \right] .		\label{eq:ehlersj3}
\eeq
Note that we are using $\tilde{j}$ to denote the Noether current for Ehlers symmetry when viewed as a symmetry of $I_{\rm 2d}$, and we are using $\hat{j}$ to denote the Noether current for Ehlers symmetry when viewed as a symmetry of $\hat{I}_{\rm 2d}$.  
The current $\tilde{j}_3$ is almost the same as the corresponding Matzner-Misner current.  We recall that the latter is given by (c.f. eqn. \ref{eq:noether3})
\beq
j_3	=  \frac{\beta}{4} \left[ -2 \phi \partial U
				- \phi^{-1} e^{4U} A \partial A 
				+ \partial \phi \right] .		\label{eq:noether3}
\eeq
Integrating the currents to get the charges gives
\beq\label{eq:level}
\tilde{q}_3 = q_3 - \frac{A_H}{4} \,.
\eeq
The last term is a shift by the Bekenstein-Hawking entropy.  
In terms of the usual Kac-Moody generators, the Matzner-Misner algebra appears at degree 0 and the Ehlers algebra appears at degree 1 \cite{breitenlohner2005explicit,Breitenlohner:1986um}.  
The shift in eqn. \eqref{eq:level} is the Kac-Moody level \cite{kac1990infinite}, so $k = A_H/4$: the Kac-Moody level is the Bekenstein-Hawking entropy.

\bibliographystyle{JHEP}
\bibliography{level.bib}

\end{document}